\documentclass[a4paper,11pt]{article}
\usepackage{graphicx}
\usepackage{jheppub}
\usepackage[T1]{fontenc}
\usepackage{enumitem}
\usepackage{xcolor}
\usepackage[utf8]{inputenc}
\usepackage{bbold}
\usepackage{cleveref}
\usepackage{amsmath}
\usepackage{appendix}
\usepackage{amssymb}
\usepackage{pifont}
\usepackage{braket}
\usepackage{bm}
\usepackage{makecell}
\usepackage{subfig}
\usepackage{tabularx,booktabs}
\newcolumntype{C}{>{\centering\arraybackslash}X}
\newcommand{\sm}[1]{{\color{purple} #1}}

\usepackage{multirow}
\usepackage[normalem]{ulem}
\usepackage{comment}
\usepackage{slashed}
\usepackage[table, svgnames, dvipsnames]{xcolor}
\usepackage{hyphenat}
\usepackage{multirow,bigdelim}
\usepackage{lipsum}
\usepackage{makecell}

\newcommand{\thetanu}{\theta^{\nu}}
\newcommand{\thetal}{\theta^{l}}
\newcommand{\Ul}{U_l}
\newcommand{\Unu}{U_\nu}
\newcommand{\UPMNS}{U_{\rm PMNS}}

\hypersetup{
    colorlinks=true,
    urlcolor=emerald,anchorcolor=emerald,citecolor=emerald,
    filecolor=emerald,linkcolor=emerald,menucolor=emerald,
    pagecolor=emerald,linktocpage=true,pdfproducer=medialab
}
\definecolor{brickred}{cmyk}{0, 0.9, 0.98, 0.38}
\definecolor{emerald}{cmyk}{0.91, 0.25, 0.9, 0.2}

\title{\boldmath 
\color{brickred}{Constraining the Neutrino Mixing Matrix via Single-Sector Charged-Lepton Rotations in the JUNO Precision Era}
}
\author[1]{A.~Giarnetti,}
\author[2,3]{S.~Marciano}
\author[4]{and D.~Meloni}
\affiliation[1]{INFN Sezione di Roma, Piazzale Aldo Moro 2, 00185,
Roma, Italy}
\affiliation[2]{Instituto de F\'isica Corpuscular (CSIC-Universitat de Val\`encia),
Parc Cient\'ific UV, C/Catedr\'atico Jos\'e Beltr\'an, 2, E-46980 Paterna, Spain}
\affiliation[3]{Departament de F\'isica Te\`orica, Universitat de Val\`encia, 46100 Burjassot, Spain}
\affiliation[4]{Dipartimento di Matematica e Fisica, Universit\'a di Roma Tre, via della Vasca Navale 84, 00146 Rome, Italy}
\emailAdd{agiarnetti@roma1.infn.it}
\emailAdd{simone.marciano@ific.uv.es}
\emailAdd{davide.meloni@uniroma3.it}

\abstract{
The unprecedented precision now being achieved in the measurement of the
Pontecorvo--Maki--Nakagawa--Sakata (PMNS) lepton mixing matrix opens a
new window onto the underlying structure of the neutrino mass matrix and
the possibly associated flavor symmetries. In this work, we investigate the constraints
imposed on the unitary matrix $\Unu$ that diagonalises the neutrino mass matrix,
under the hypothesis that the charged-lepton mixing matrix $\Ul$ consists of a
single two-by-two rotation in one of the three sectors: (1,2), (1,3), or (2,3).
For this analysis, we considered the latest global fit which incorporates the precision measurement of $\theta_{12}$ from the JUNO
experiment.
For each scenario, we also derive analytical expressions for the entries of $\Unu$ in
terms of the measured PMNS parameters to obtain compact sum-rule-like formulae.
}


\begin{document}
\maketitle\newpage

\section{Introduction}
\label{sec:intro}

The discovery of neutrino oscillations confirms that neutrinos are massive and
that lepton flavor is not conserved in propagation, and stands as one of the landmark
results in modern particle physics~\cite{Fukuda:1998mi,Ahmad:2002jz,Eguchi:2002dm}.
The phenomenon is fully described in the three-flavor framework by the
Pontecorvo--Maki--Nakagawa--Sakata (PMNS) mixing
matrix~\cite{Pontecorvo:1957qd,Pontecorvo:1967fh,Maki:1962mu}, a $3\times3$ unitary
matrix conventionally parameterized in terms of three mixing angles
$\theta_{12}$, $\theta_{13}$, $\theta_{23}$, one Dirac CP-violating phase $\delta_{\rm CP}$,
and, for Majorana neutrinos, two additional Majorana phases. Over the past two decades,
a global experimental program has determined all three mixing angles with progressively
increasing precision~\cite{Esteban:2024asa,nufit61,ParticleDataGroup:2024cfk}.

\medskip
The field is now entering a genuine \textit{precision era}. The reactor experiments
Daya~Bay~\cite{DayaBay:2022orm}, RENO~\cite{RENO:2018dro}, and
Double~Chooz~\cite{DoubleChooz:2019qbj} have measured $\theta_{13}$ with sub-percent
precision, while the long-baseline oscillation experiments T2K~\cite{T2K:2023smv} and
NOvA~\cite{NOvA:2021nfi} have significantly sharpened the determination of $\theta_{23}$
and provided first hints on the CP phase and the mass ordering. A major milestone was
recently reached by the Jiangmen Underground Neutrino Observatory (JUNO), which reported
its first measurement of the solar oscillation parameters from  59.1 days of
data taking~\cite{JUNO:2025theta12}:
\begin{equation}
  \sin^2\theta_{12} = 0.3092 \pm 0.0087 \,.
  \label{eq:juno_th12}
\end{equation}
This result improves the combined precision of all previous measurements of $\theta_{12}$ by a
factor of 1.6 and inspired a large number of study regarding the implication of such a precise measurement of the solar mixing angle \cite{Petcov:2025aci,Dutta:2026dzh,Nanda:2025fvw,Kumar:2025lkv,Borah:2025vtn,Ding:2025dqd,Xing:2025bdm,Zhang:2025jnn,He:2025idv,Ge:2025csr,Chen:2025afg,Jiang:2025hvq,Ge:2025cky,Huang:2025znh,Xing:2025xte,Chao:2025sao}. Notice that, with the full JUNO data set, the precision on $\theta_{12}$ is expected
to reach the sub-percent level~\cite{JUNO:2015zny}.

\medskip
Looking further ahead, the Deep Underground Neutrino Experiment
(DUNE)~\cite{DUNE:2020ypp,DUNE:2020jqi} and the Tokai-to-Hyper-Kamiokande experiment
(T2HK)~\cite{Hyper-Kamiokande:2018ofw} will reach a better precision on $\theta_{23}$, $\delta_{\rm CP}$,
and the neutrino mass ordering. In particular, DUNE is
designed to measure $\sin^2\theta_{23}$ to better than 1\% and to determine
$\delta_{\rm CP}$ with a sensitivity exceeding $5\sigma$ for a wide range of true values.
These improvements will essentially complete the picture of the PMNS matrix, fixing all of
its parameters with high accuracy and rendering the PMNS matrix a precision testing ground
for theories of flavor.

\medskip
This prospect motivates a renewed interest in the possibly hidden theoretical structures that could
underlie the observed pattern of lepton mixing. It has long been recognized that the
PMNS matrix arises as the mismatch between the unitary transformations that diagonalize
the charged-lepton and the neutrino mass matrices~\cite{Schechter:1980gr}.
Writing the relation as:
\begin{equation}
  \UPMNS = \Ul\, \Unu^\dagger \,,
  \label{eq:pmns_factorization}
\end{equation}
where $\Ul$ ($\Unu$) is the unitary matrix arising from the diagonalization of the
charged-lepton (neutrino) mass matrix, the observed PMNS matrix encodes the combined
effect of \textit{both} sectors. A large body of work has explored the hypothesis that
the mixing pattern originates primarily from the neutrino
sector~\cite{Harrison:1999cf,Harrison:2002er,Xing:2002sw,Vissani:1997pa,
Altarelli:2005yx,King:2011ab,Giarnetti:2025idu,Marciano:2024nwm,Barger:2001yr} with
subleading corrections from the charged-lepton sector.

\medskip
A particularly clean and predictive framework is to assume that $\Ul$ is restricted to a
single rotation in one of the three $(i,j)$ planes, parameterized by a single angle
$\theta_l$. Such an assumption is motivated by:
\begin{itemize}
  \item \textbf{CKM analogy}: In the quark sector, the mixing is hierarchical and
    dominated by the Cabibbo angle in the (1,2) plane. Models of quark-lepton
    complementarity or unified flavor symmetries often predict analogous structures in
    the charged-lepton sector~\cite{Antusch:2007rk,Giarnetti:2024vgs}.
  \item \textbf{Residual symmetries}: Discrete non-Abelian flavor symmetries~\cite{Altarelli:2010gt,King:2013eh,Feruglio:2019ybq},
    such as $A_4$, $S_4$, or $\Delta(27)$, can be broken to different residual
    subgroups in the charged-lepton and neutrino sectors, often leading to a
    block-diagonal structure in $\Ul$ where only one rotation is present at leading order.
  \item \textbf{Predictivity}: Imposing a single-sector structure on $\Ul$ allows
    one to easily invert the relation~\eqref{eq:pmns_factorization} and extract the
    neutrino matrix $\Unu = \UPMNS^\dagger\,\Ul$ in terms of measured quantities,
    yielding testable predictions on possible symmetries in the $\Unu$ structure.
\end{itemize}

\noindent The idea was first explored in detail in the (1,2) sector in~\cite{Antusch:2007rk} 
and applied to test tri-bimaximal and bimaximal predictions for the
neutrino mixing angle $\thetanu_{12}$. With the much more precise data now available
and forthcoming, sum rules obtained in this framework may become powerful discriminators among flavor models.

\medskip
In this paper, we revisit and extend this analysis to all three single-rotation
hypotheses for $\Ul$, derive the analytical form of $\Unu$ in each case, and discuss
the impact of the current precision data---particularly JUNO's measurement of
$\theta_{12}$---on the allowed structure of the neutrino mass matrix. The paper is
organized as follows. Section~\ref{sec:formalism} develops the general formalism,
introduces the parameterization of the PMNS matrix, and derives the analytical
expressions for $\Unu$ in the three cases together with approximate sum rules valid
to leading order in $\theta_{13}$. Sections~\ref{sec:results_12},
\ref{sec:results_13}, and~\ref{sec:results_23} present the numerical constraints
on the entries of $\Unu$ in each scenario, while in \ref{sec:deltaCP} and \ref{sec:jarlskog} we investigate the effect of $\delta_{\rm{CP}}$ on the neutrino angle measurements and the possible values of the Jarlskog invariant of the neutrino matrix, respectively. Section~\ref{sec:conclusions} contains
our conclusions.

\section{Framework and Formalism}
\label{sec:formalism}

\subsection{The standard parametrization of the PMNS mixing matrix}
\label{ssec:pmns_param}

In the Standard Model extended with massive neutrinos, the charged-current Lagrangian
in the mass eigenstate basis contains the lepton mixing matrix $U_\text{PMNS}$, that depends on the neutrino and charged lepton mixing matrices, as in Eq.~\eqref{eq:pmns_factorization}. We adopt the convention in Eq.~\eqref{eq:pmns_factorization} (i.e. $\UPMNS = \Ul\, \Unu^\dagger$),
where $\Ul$ and $\Unu$ are $3\times3$ unitary matrices that bring the charged-lepton
and neutrino mass matrices to diagonal form, respectively. In the common convention
where one works in the basis where the charged leptons are diagonal ($\Ul = \mathbb{1}$),
the full mixing is attributed to $\Unu$, i.e.\ $\UPMNS = \Unu^\dagger$. Beyond this
special basis, both $\Ul$ and $\Unu$ contribute non-trivially.

\medskip
The PDG parameterization~\cite{ParticleDataGroup:2024cfk} writes the PMNS matrix as
\begin{equation}
  \UPMNS = R_{23}(\theta_{23})\,U_{13}(\theta_{13},\delta_{\rm CP})\,R_{12}(\theta_{12}){\cdot\mathcal{P}}\,,
  \label{eq:upmns_param}
\end{equation}
where $R_{ij}(\theta)$ denotes a real rotation by an angle $\theta$ in the $(i,j)$ plane,
and $U_{13}(\theta_{13},\delta_{\rm CP})$ is the complex rotation
\begin{equation}
  U_{13}(\theta_{13},\delta_{\rm CP})
  = \begin{pmatrix} c_{13} & 0 & s_{13}\,e^{-i\delta_{\rm CP}} \\
                    0 & 1 & 0 \\
                    -s_{13}\,e^{i\delta_{\rm CP}} & 0 & c_{13}
    \end{pmatrix}\,.
\end{equation}
Here $c_{ij} \equiv \cos\theta_{ij}$, $s_{ij}\equiv \sin\theta_{ij}$, with
$\theta_{ij}\in[0,\pi/2]$ and $\delta_{\rm CP}\in[0,2\pi)$. Majorana phases are
contained in a diagonal matrix $\mathcal{P} = \mathrm{diag}(e^{i\alpha_1},e^{i\alpha_2},1)$
that does not affect oscillation probabilities and is omitted hereafter. The explicit
PMNS matrix reads :
\begin{equation}
  \UPMNS =
  \begin{pmatrix}
    c_{12}c_{13} & s_{12}c_{13} & s_{13}e^{-i\delta} \\
    -s_{12}c_{23}-c_{12}s_{23}s_{13}e^{i\delta} &
     c_{12}c_{23}-s_{12}s_{23}s_{13}e^{i\delta} &
     s_{23}c_{13} \\
     s_{12}s_{23}-c_{12}c_{23}s_{13}e^{i\delta} &
    -c_{12}s_{23}-s_{12}c_{23}s_{13}e^{i\delta} &
     c_{23}c_{13}
  \end{pmatrix}\,,
  \label{eq:upmns_explicit}
\end{equation}
where we suppress the subscript on $\delta_{\rm CP}$ for brevity.
The three PMNS mixing angles can be extracted from the matrix elements via
\begin{equation}
  \sin^2\theta_{13} = |U_{e3}|^2, \qquad
  \sin^2\theta_{12} = \frac{|U_{e2}|^2}{1-|U_{e3}|^2}, \qquad
  \sin^2\theta_{23} = \frac{|U_{\mu3}|^2}{1-|U_{e3}|^2}\,.
  \label{eq:angle_extraction}
\end{equation}
The same prescription applies to extract the mixing angles of any unitary matrix
once it has been brought to the standard parametrisation in Eq.~\eqref{eq:upmns_param} via rephasing of rows and columns.

\subsection{Inverting the factorization}
\label{ssec:inversion}
Given a hypothesis for $\Ul$, the neutrino mixing matrix is obtained by inverting
Eq.~\eqref{eq:pmns_factorization}:
\begin{equation}
   {\Unu = \UPMNS^\dagger\,\Ul\,.}
  \label{eq:Unu_formula}
\end{equation}
The matrix $\Unu$ is then brought to the PDG form~\eqref{eq:upmns_param} after
absorbing unphysical phases, and its mixing angles $\thetanu_{12}$, $\thetanu_{13}$,
$\thetanu_{23}$ and CP phase $\delta^\nu$ are extracted via
Eq.~\eqref{eq:angle_extraction} applied to $\Unu$.

\subsection{Single-sector charged-lepton rotations}
\label{ssec:single_rotations}

We consider three distinct hypotheses for the structure of $\Ul$\footnote{Notice that the considered $\Ul$ structures which only contain one leptonic mixing angle do not allow the presence of observable phases.}:

\medskip
\noindent\textbf{Case I -- (1,2) rotation:}
\begin{equation}
  \Ul^{(12)} = R_{12}(\thetal_{12}) =
  \begin{pmatrix}
    \cos\thetal_{12} & \sin\thetal_{12} & 0 \\
    -\sin\thetal_{12} & \cos\thetal_{12} & 0 \\
    0 & 0 & 1
  \end{pmatrix}\,.
  \label{eq:Ul_12}
\end{equation}

\noindent\textbf{Case II -- (1,3) rotation:}
\begin{equation}
  \Ul^{(13)} = R_{13}(\thetal_{13}) =
  \begin{pmatrix}
    \cos\thetal_{13} & 0 & \sin\thetal_{13} \\
    0 & 1 & 0 \\
    -\sin\thetal_{13} & 0 & \cos\thetal_{13}
  \end{pmatrix}\,.
  \label{eq:Ul_13}
\end{equation}

\noindent\textbf{Case III -- (2,3) rotation:}
\begin{equation}
  \Ul^{(23)} = R_{23}(\thetal_{23}) =
  \begin{pmatrix}
    1 & 0 & 0 \\
    0 & \cos\thetal_{23} & \sin\thetal_{23} \\
    0 & -\sin\thetal_{23} & \cos\thetal_{23}
  \end{pmatrix}\,.
  \label{eq:Ul_23}
\end{equation}

\medskip
Each of these can be motivated by different classes of flavor models. The (1,2)
rotation is the most widely studied and is directly inspired by the CKM
matrix~\cite{Cabibbo:1963yz,Kobayashi:1973fv}, in which the dominant mixing arises
from the Cabibbo angle $\theta_C \approx 13^\circ$ in the (1,2)
plane~\cite{Antusch:2007rk}. This structure naturally arises in
grand unified theories with $SU(5)$ and $SO(10)$ symmetries, where the charged-lepton Yukawa matrix
is related to the down-quark Yukawa matrix and thus inherits a similar hierarchical
texture~\cite{Georgi:1974sy}.
The (2,3) rotation can be motivated by models with $\mu$--$\tau$
symmetry~\cite{Arcadi:2022ojj,Ma:2001mr,Mohapatra:2004mf}, which predicts a
maximal atmospheric angle $\theta_{23} = \pi/4$ and $\theta_{13} = 0$ in the neutrino
sector, with the observed $\theta_{23}$ close to maximal receiving corrections from the
charged-lepton sector through a (2,3) rotation. The (1,3) rotation, while less commonly
studied, appears in models where leptonic mixing in the first and third
generation is generated at leading order, and can arise in certain Froggatt--Nielsen
mechanisms~\cite{Froggatt:1978nt} or specific left-right symmetric
models~\cite{Pati:1974yy,Mohapatra:1974gc}.

\subsection{Analytical expressions for \boldmath{$\Unu$} and sum-rules}
\label{ssec:analytical}

In each case, $\Unu = \UPMNS^\dagger\,\Ul$ is computed by multiplying $\UPMNS^\dagger$
on the right by a single rotation. Since $R_{ij}$ mixes only columns $i$ and $j$,
the remaining column of $\UPMNS^\dagger$ is left untouched. This leads to a key
structural observation: \textit{the column of $\Unu$ whose index does not appear in
the $(i,j)$ sector of $\Ul$ is identical to the corresponding column of $\UPMNS^\dagger$,
and is therefore entirely determined by the PMNS parameters, independently of $\thetal$.}
Specifically: in Case~I the \textbf{third} column is preserved; in Case~II the
\textbf{second} column while in Case~III is the \textbf{first} column which is left untouched. 
A few comments on the notation are in order.
Since some entries of the matrix $\Unu$ are not affected by the single--sector
rotation in the charged--lepton sector, we display explicitly only the portion
of $\Unu$ that is modified by $\Ul$. The elements that coincide with those of
$\UPMNS^\dagger$ are left implicit in order to avoid cluttering the expressions.
We denote the entries of $\UPMNS$ as $u_{ij}$, which can be directly read from
Eq.~\eqref{eq:upmns_explicit}. Furthermore, we use the shorthand notation
$c_l \equiv \cos\theta^l_{ij}$ and $s_l \equiv \sin\theta^l_{ij}$ for the cosine
and sine appearing in the single--sector rotation matrix
$\Ul = R_{ij}(\theta^l_{ij})$.

Before examining the specific rotation cases, we mention here that applying~\eqref{eq:angle_extraction} in the limit of small $\theta_{13}$ and $\theta_{ij}^l$,  we obtain the approximate relations:
\begin{align}
\sin^2\thetanu_{13} &\approx
  \sin^2\theta_{12}\sin^2\theta_{23}
 \,,\label{eq:th13nu_LO_13b}\\
\sin^2\thetanu_{23} &\approx
  \frac{\cos^2\theta_{12}\sin^2\theta_{23}}{1 - \sin^2\theta_{12}\sin^2\theta_{23}}
  \,,
  \label{eq:th23nu_LO_13}\\
  \sin^2\thetanu_{12} &\approx
\frac{\sin^2\theta_{12}\cos^2\theta_{23}}{1 - \sin^2\theta_{12}\sin^2\theta_{23}}
 \,.
  \label{eq:th12nu_LO_13}
\end{align}
The preceding relations demonstrate that in the limit $\theta_{13}, \theta_{ij} \to 0$, a clear hierarchy among the mixing angles in $U_{\nu}$ emerges. This hierarchy is entirely independent of the specific rotation angle of the charged leptons:
\begin{equation}
    \sin^2 \theta_{23}^{\nu} > \sin^2 \theta_{13}^{\nu} \sim \sin^2 \theta_{12}^{\nu} \,.
\end{equation}
The latter two mixing angles take similar values; which of the two is larger depends on the next-to-leading order $\theta_{13}$ corrections and the specific octant of $\theta_{23}$. Finally, we notice in this limit the following relation:
\begin{equation}
\label{eq:caseIIhierarchy}
\dfrac{\sin^2\theta_{13}^\nu \sin^2\theta_{12}^\nu}{\sin^2\theta_{23}^\nu}\approx \tan^2\theta_{12}\sin^2\theta_{12}\cos^2\theta_{23}\,.
\end{equation}

\subsubsection{Case I: \boldmath{$(1,2)$} sector rotation}
\label{sssec:case_12}

With $\Ul = R_{12}(\thetal_{12})$, the matrix $\Unu^{(12)} = \UPMNS^\dagger\,R_{12}(\thetal_{12})$
reads:
\begin{equation}
  U_\nu^{(12)} =
  \begin{pmatrix}
     c_l c_{12}c_{13} + s_l\mathcal{K}_1 &
    s_l c_{12}c_{13} - c_l\mathcal{K}_1 & u_{31}^\ast\\[6pt]
    c_l s_{12}c_{13} - s_l\mathcal{K}_2 &
    s_l s_{12}c_{13} + c_l\mathcal{K}_2 &u_{32}^\ast \\[6pt]
    c_l s_{13}e^{i\delta} - s_l s_{23}c_{13} &
    s_l s_{13}e^{i\delta} + c_l s_{23}c_{13} &u_{33}^\ast
  \end{pmatrix}\,,
  \label{eq:Unu_12_explicit}
\end{equation}
with 
\begin{equation}\label{eq:K1andK2}
\begin{cases}
    \mathcal{K}_1=s_{12}c_{23}+c_{12}s_{23}s_{13}e^{-i\delta}\\[6pt]
    \mathcal{K}_2=c_{12}c_{23}-s_{12}s_{23}s_{13}e^{-i\delta}
    \end{cases}\,.
\end{equation}
\medskip
Applying Eq.~\eqref{eq:angle_extraction} to the $\thetal_{12}$-independent third
column, we obtain the same result mentioned in Eq. (\ref{eq:th13nu_LO_13b}) which,  in this case, is exact in the limit $\theta_{13}\to0$ and not valid only for small $\thetal_{12}$. 
Expanding to the next to leading leading order in the small reactor angle $\sin\theta_{13}$ one obtain:
\begin{equation}
   {
  \sin^2\thetanu_{13} \approx \sin^2\theta_{12}\sin^2\theta_{23}
  - \tfrac{1}{2}\sin 2\theta_{12}\,\sin 2\theta_{23}\sin\theta_{13}\cos\delta
  + \mathcal{O}(\sin^2\theta_{13})\,.}
  \label{eq:th13nu_LO_12}
\end{equation}
The leading term is fixed entirely by the solar and atmospheric PMNS angles, with
the $\mathcal{O}(\sin\theta_{13})$ correction sensitive to the CP phase. The (2,3) and (3,3) entries of Eq. (\ref{eq:Unu_12_explicit})
 are again independent from the leptonic rotation and Eq. (\ref{eq:th23nu_LO_13}) holds for all $\thetal_{12}$. 
Thus, in this particular leptonic rotation there are two testable
predictions independent of the unknown $\thetal_{12}$: any flavor model specifying
$\thetanu_{13}$ or $\thetanu_{23}$ can be confronted with measured PMNS angles without
any additional free parameter. Note in particular that since $\sin^2\theta_{12} \sim 0.3$,
one finds $\sin^2\thetanu_{23} > \sin^2\theta_{23}$ at LO: the neutrino atmospheric
angle is predicted to be larger than its PMNS counterpart.

The angle $\thetanu_{12}$ does depend on $\thetal_{12}$ through the first two columns.
At leading order in $\sin\theta_{13}$ and small $\thetal_{12}$, the (1,1) and (1,2) entries imply:
\begin{equation}
  \thetanu_{12} \approx \theta_{12} - \thetal_{12}\,\cos\theta_{23}
  + \mathcal{O}({{\theta^{l^2}_{12}}},\,\sin\theta_{13})\,,
  \label{eq:th12nu_sumrule_12}
\end{equation}
which generalizes the sum rule of Ref.~\cite{Antusch:2007rk} that was obtained with a different parameterization of the $\Unu$ matrix.

\subsubsection{Case II: \boldmath{$(1,3)$} sector rotation}
\label{sssec:case_13}

With $\Ul = R_{13}(\thetal_{13})$, the matrix $\Unu^{(13)} = \UPMNS^\dagger\,R_{13}(\thetal_{13})$
reads:
\begin{equation}
  \Unu^{(13)} =
  \begin{pmatrix}
    c_l c_{12}c_{13} - s_l\mathcal{K}_3 &
   u^\ast_{21} &
    s_l c_{12}c_{13} + c_l\mathcal{K}_3 \\[6pt]
    c_l s_{12}c_{13} + s_l\mathcal{K}_4 &
    u^\ast_{22} &
    s_l s_{12}c_{13} - c_l\mathcal{K}_4 \\[6pt]
    c_l s_{13}e^{i\delta} - s_l c_{23}c_{13} &
    u^\ast_{23} &
    s_l s_{13}e^{i\delta} + c_l c_{23}c_{13}
  \end{pmatrix}\,,
  \label{eq:Unu_13_explicit}
\end{equation}
with
\begin{equation}\label{eq:K3andK4}
    \begin{cases}
        \mathcal{K}_3=\,s_{12}s_{23}-c_{12}c_{23}s_{13}e^{-i\delta}\\[6pt]
        \mathcal{K}_4=c_{12}s_{23}+s_{12}c_{23}s_{13}e^{-i\delta}
    \end{cases}\,.
\end{equation}
\medskip

In this case no compact additional relations independent of the rotation angle $\thetal_{13}$ can be easily obtained. 

\subsubsection{Case III: \boldmath{$(2,3)$} sector rotation}
\label{sssec:case_23}

With $\Ul = R_{23}(\thetal_{23})$, making use of the Eqs.~\eqref{eq:K1andK2} and~\eqref{eq:K3andK4}, we obtain:
    \begin{equation}
\small
  \Unu^{(23)} =
  \begin{pmatrix}
    c_{12}c_{13} &
    -c_l\mathcal{K}_1 - s_l\mathcal{K}_3 &
     s_l\mathcal{K}_1 - c_l\mathcal{K}_3 \\[6pt]
    s_{12}c_{13} &
    c_l\mathcal{K}_2 + s_l\mathcal{K}_4 &
   -s_l\mathcal{K}_2 + c_l\mathcal{K}_4 \\[6pt]
    s_{13}e^{i\delta} &
    c_l s_{23}c_{13} - s_l c_{23}c_{13} &
    s_l s_{23}c_{13} + c_l c_{23}c_{13}
  \end{pmatrix}\,.
  \label{eq:Unu_23_explicit}
\end{equation}

\medskip
Using the angle-subtraction identities, the third-row entries simplify exactly to
$(\Unu^{(23)})_{32} = \cos\theta_{13}\sin(\theta_{23}-\thetal_{23})$ and
$(\Unu^{(23)})_{33} = \cos\theta_{13}\cos(\theta_{23}-\thetal_{23})$.
Thus, in the limit of small $\theta_{13}$,
the full matrix reduces to:
\begin{equation}
\small
  \Unu^{(23)} \approx
  \begin{pmatrix}
    \cos\theta_{12} &
    -\sin\theta_{12}\cos(\theta_{23}-\thetal_{23}) &
     \sin\theta_{12}\sin(\theta_{23}-\thetal_{23}) \\[4pt]
    \sin\theta_{12} &
     \cos\theta_{12}\cos(\theta_{23}-\thetal_{23}) &
    -\cos\theta_{12}\sin(\theta_{23}-\thetal_{23}) \\[4pt]
    0 &
    -\sin(\theta_{23}-\thetal_{23}) &
     \cos(\theta_{23}-\thetal_{23})
  \end{pmatrix}
  = R_{12}(\theta_{12})\,R_{23}(\theta_{23}-\thetal_{23})\,.
  \label{eq:Unu_23_LO}
\end{equation}
This is a product of two real rotations. Crucially, the ordering of this product is $R_{12}\,R_{23}$,
which differs from the PDG convention $R_{23}\,U_{13}\,R_{12}$; thus one cannot simply
read off $\thetanu_{13} = 0$ from the block structure in this approximation. Applying the standard
formula~\eqref{eq:angle_extraction} entry by entry, one finds:
\begin{equation}
  \sin^2\thetanu_{13}
  \approx \sin^2\theta_{12}\,\sin^2(\theta_{23}-\thetal_{23})\,,
  \label{eq:th13nu_case3_LO}
\end{equation}
\begin{equation}
   {
  \sin^2\thetanu_{12}
  \approx
  \frac{\sin^2\theta_{12}\cos^2(\theta_{23}-\thetal_{23})}
       {1 - \sin^2\theta_{12}\sin^2(\theta_{23}-\thetal_{23})}\,,}
  \label{eq:th12nu_case3_LO}
\end{equation}
\begin{equation}
   {
  \sin^2\thetanu_{23}
  \approx
  \frac{\cos^2\theta_{12}\sin^2(\theta_{23}-\thetal_{23})}
       {1 - \sin^2\theta_{12}\sin^2(\theta_{23}-\thetal_{23})}\,.}
  \label{eq:th23nu_case3_LO}
\end{equation}
All three neutrino angles depend on the atmospheric sector only through the single
effective combination $\Delta_{23} \equiv \theta_{23}-\thetal_{23}$. Note that
$\thetanu_{13}$ is generically nonzero,  being vanishing only in the degenerate case
$\theta_{23} = \thetal_{23}$ or for in the unphysical limit $\theta_{12} = 0$.

For small $\thetal_{23}$, expanding to linear order recovers compact approximate
expressions:
\begin{align}
  \thetanu_{12} &\approx \theta_{12} + \mathcal{O}({\theta^{l^2}_{23}})\,,
  \label{eq:th12nu_case3_small}\\
  \thetanu_{23} &\approx \theta_{23} - \thetal_{23} + \mathcal{O}({\theta^{l^2}_{23}})\,,
  \label{eq:th23nu_case3_small}
\end{align}
giving the compact atmospheric sum rule
\begin{equation}
   {\thetanu_{23} \approx \theta_{23} - \thetal_{23}
  + \mathcal{O}({\theta^{l^2}_{23}},\theta_{13}^2)\,.}
  \label{eq:sumrule_23}
\end{equation}
Going beyond the vanishing $\theta_{13}$ limit, one can note that the (1,1)
entry equals $\cos\theta_{12}\cos\theta_{13}$ exactly, and in the PDG parameterization the (1,1) entry
of $\Unu$ is $\cos\thetanu_{12}\cos\thetanu_{13}$. Equating the two gives the exact,
$\thetal_{23}$-independent sum-rule:
\begin{equation}
   {
  \cos^2\thetanu_{12}\,\cos^2\thetanu_{13} = \cos^2\theta_{12}\,\cos^2\theta_{13}\,.}
  \label{eq:sumrule_case3_exact}
\end{equation}
This holds for any value of $\thetal_{23}$, any PMNS parameters, and without any
expansion in $\theta_{13}$. It constrains the allowed values of $\thetanu_{12}$ and
$\thetanu_{13}$ to lie on the curve defined by Eq.~\eqref{eq:sumrule_case3_exact}
in the $(\thetanu_{12},\thetanu_{13})$ plane. For instance, a model predicting
$\thetanu_{13} = 0$ forces $\cos^2\thetanu_{12} = \cos^2\theta_{12}\cos^2\theta_{13}$,
i.e.\ $\thetanu_{12} > \theta_{12}$. Conversely, a model predicting
$\thetanu_{12} = \theta_{12}$ forces $\thetanu_{13} = \theta_{13}$ exactly.
At leading order in the small reactor angle $\theta_{13}$,
Eq.~\eqref{eq:sumrule_case3_exact} reduces to $\thetanu_{12} \approx \theta_{12}$,
consistent with the leading-order result obtained with $\theta_{13}=0$.

\section{Constraints on \boldmath{$\Unu$}}
\label{sec:results}

\begin{table}[tbp]
\centering
\renewcommand{\arraystretch}{1.6}
\begin{tabular}{lcc}
\toprule
\bf{Parameter}
  & \bf{NuFit~6.0 (w/o JUNO)}
  & \bf{NuFit~6.1 (with JUNO)} \\
\midrule
$\sin^2\theta_{12}$
  & $0.307 \pm 0.012$
  & $0.3088 \pm 0.0067$ \\
$\sin^2\theta_{13}$
  & $0.0222 \pm 0.0006$
  & $0.0225 \pm 0.0006$ \\
$\sin^2\theta_{23}$ (\textbf{LO})
  & $0.470 \pm 0.015$
  & $0.470 \pm 0.015$ \\
$\sin^2\theta_{23}$ (\textbf{UO})
  & $0.561 \pm 0.013$
  & $0.561 \pm 0.013$ \\
$\delta_{\rm CP}$
  & $212^\circ$
  & $212^\circ$ \\
\bottomrule
\end{tabular}
\caption{PMNS parameters for Normal Ordering used in this analysis, from the
NuFit~6.0 (without JUNO)~\cite{Esteban:2024asa} and NuFit~6.1 (with JUNO)~\cite{nufit61}
global fits, both including SK atmospheric data. We report best-fit values with
$1\sigma$ uncertainties. For $\theta_{23}$, both the lower octant (\textbf{LO}) and upper
octant (\textbf{UO}) solutions are shown; the \textbf{UO} value is taken as the reflection of the \textbf{LO}
minimum used in the scan (see text). When the published $1\sigma$ interval is asymmetric, we symmetrize by averaging the upper and lower uncertainties.}
\label{tab:nufit}
\end{table}

\noindent
For our numerical analysis which in this section will focus on the constraints on the neutrino mixing angles, we use the PMNS mixing angles from the NuFit global
fits~\cite{Esteban:2024asa,nufit61}. To highlight the impact of the JUNO measurement, we
compare two sets of input parameters:
\begin{itemize}
  \item \textbf{NuFit~6.1} (with JUNO)~\cite{nufit61}: the November~2025 update
    including the JUNO $\theta_{12}$ result \cite{JUNO:2025theta12}, with SK atmospheric data;
  \item \textbf{NuFit~6.0} (without JUNO)~\cite{Esteban:2024asa}: the September~2024
    release prior to the JUNO measurement, also with SK atmospheric data.
\end{itemize}
Both sets of best-fit values and $1\sigma$ uncertainties for \textbf{NO} are collected in
Table~\ref{tab:nufit}. For the upper octant (\textbf{UO}) of $\theta_{23}$, the best-fit
value is taken as the reflection of the lower-octant (\textbf{LO}) minimum about
$\sin^2\theta_{23}=0.5$, i.e.\ $\sin^2\theta_{23}^{\rm UO} = 1 - \sin^2\theta_{23}^{\rm LO} + \epsilon$
with a small offset consistent with the NuFit $\chi^2$ profile. Throughout this work
we assume Normal Ordering (\textbf{NO}). The CP phase
$\delta_{\rm CP}$ is scanned over its full range $[0, 2\pi)$ in the following numerical analysis.
\begin{figure}[tb]
    \centering
    \includegraphics[width=0.7\linewidth]{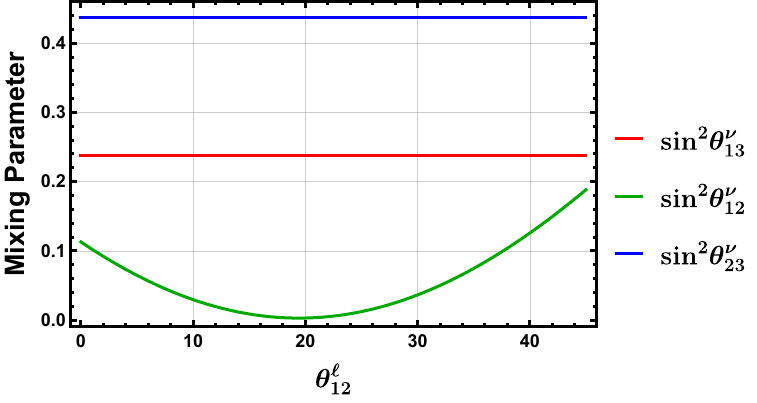}
    \caption{Neutrino mixing angles $\sin^2\thetanu_{13}$ (red),
    $\sin^2\thetanu_{12}$ (green), and $\sin^2\thetanu_{23}$ (blue)
    as functions of $\thetal_{12}$, with the PMNS angles fixed to
    the NuFit~6.1 best-fit values (\textbf{UO}) in \textbf{NO}.}
    \label{fig:angles_vs_theta_12}
\end{figure}
In the numerical scan, the three mixing angles are varied as Gaussian
distributions centered on the best-fit values with $1\sigma$ widths as given in
Table~\ref{tab:nufit}, while $\delta_{\rm CP}$ is drawn uniformly in $[0,2\pi)$. For
each set of PMNS inputs and a given value of the single charged-lepton rotation angle
$\thetal \in [0^\circ, 45^\circ]$ (we restrict our analyses in the lower octant of the leptonic angle for simplicity), the neutrino mixing matrix $\Unu$ is obtained via
Eqs.~\eqref{eq:Unu_12_explicit}, \eqref{eq:Unu_13_explicit},
or~\eqref{eq:Unu_23_explicit}, and the neutrino mixing angles
$\thetanu_{12}$, $\thetanu_{13}$, $\thetanu_{23}$ are extracted using Eq.~\eqref{eq:angle_extraction}.
The results are presented for each of the three
rotation cases discussed above, and compared to the analytical expectations derived in
Section~\ref{ssec:analytical}.

\subsection{Case I: (1,2) charged-lepton rotation}
\label{sec:results_12}
\begin{figure}[tbp]
    \centering
    \includegraphics[width=0.95\linewidth]{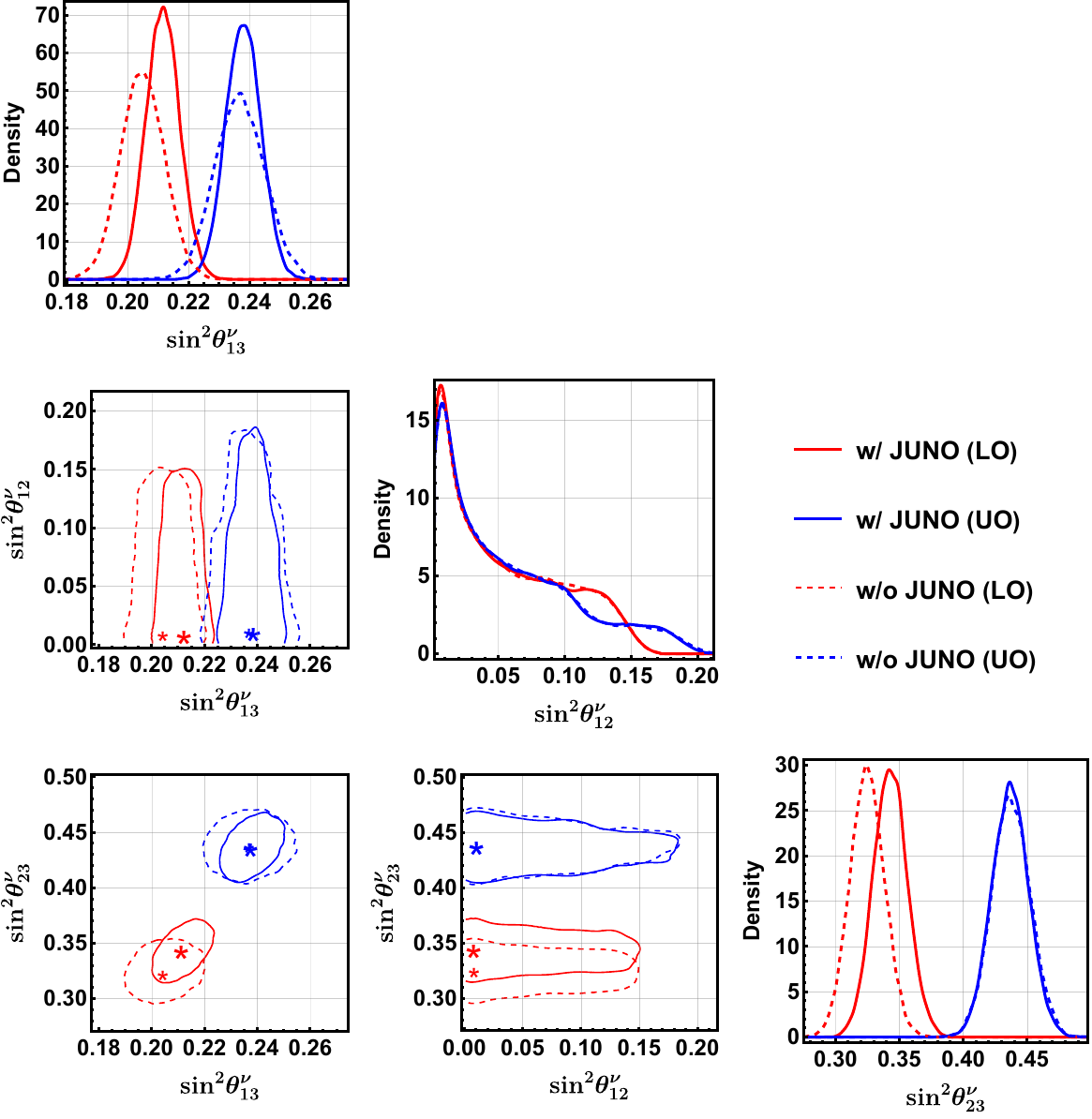}
    \caption{Distributions of the neutrino mixing angles
    $\sin^2\thetanu_{13}$, $\sin^2\thetanu_{12}$, and $\sin^2\thetanu_{23}$
    obtained by varying the PMNS angles as Gaussian distributions with $1\sigma$
    widths and scanning $\thetal_{12}\in[0^\circ,45^\circ]$ uniformly.
    The correlation plots show the regions containing 90\% of the neutrino mixing angle pairs
    $(\sin^2\thetanu_{12},\sin^2\thetanu_{23})$,
    $(\sin^2\thetanu_{13},\sin^2\thetanu_{23})$, and
    $(\sin^2\thetanu_{13},\sin^2\thetanu_{12})$,
    obtained by varying PMNS parameters and scanning $\thetal_{12}$
    over its full range. 
    Solid (dashed) curves correspond to NuFit~6.1 (NuFit~6.0) inputs; red (blue)
    curves refer to the \textbf{LO} (\textbf{UO}) solution of $\theta_{23}$.
    Stars indicate the most probable point (peak of the distribution)
    in each dataset.
    }
    \label{fig:triangle12}
\end{figure}

Figure~\ref{fig:angles_vs_theta_12} shows how the three neutrino mixing angles evolve
as a function of $\thetal_{12}$, with all PMNS parameters set to their best-fit values.
It is clear that $\sin^2\thetanu_{23}$ (blue curve) is essentially
flat: this is precisely the analytical prediction of Eq.~\eqref{eq:th23nu_LO_13},
which gives $\sin^2\thetanu_{23} \approx \cos^2\theta_{12}\sin^2\theta_{23}/(1-\sin^2\theta_{12}\sin^2\theta_{23})$,
a quantity entirely independent of $\thetal_{12}$. At the best-fit values this
evaluates to approximately 0.47, consistent with the plateau seen in the figure.
The neutrino reactor angle $\sin^2\thetanu_{13}$ (red) is also independent to the leptonic mixing angle, approximately at
$\sin^2\theta_{12}\sin^2\theta_{23} \approx 0.17$. The solar angle $\sin^2\thetanu_{12}$ (green) carries the
full dependence on $\thetal_{12}$ through the first two columns of $\Unu$, decreasing from
its PMNS value toward zero and then raising until roughly the starting value for large $\thetal_{12}$.
Figure~\ref{fig:triangle12} shows the marginal distributions of the three
neutrino mixing angles together with the 90\% contours in the three pairs of neutrino
angle planes. The contours have been obtained by varying the PMNS mixing parameters
as Gaussian distributions around each best-fit value (except $\delta_{\mathrm{CP}}$,
which is fixed to its best value) and scanning $\thetal_{12}$ uniformly in $[0^\circ,45^\circ]$;
the regions shown contain 90\% of the resulting points.
The impact of JUNO is most pronounced for $\sin^2\thetanu_{13}$:
since its leading-order value depends on $\sin^2\theta_{12}$, the improved precision on this angle from
NuFit~6.1 (cf.\ Table~\ref{tab:nufit}) translates into a slightly narrower
distribution and tighter contours (solid vs.\ dashed) in the planes involving the aforementioned angle.
The \textbf{LO} and \textbf{UO} distributions are well separated for both $\sin^2\thetanu_{13}$ and $\sin^2\thetanu_{23}$. In particular, the \textbf{UO} solution predicts larger $\sin^2\thetanu_{13}$ and $\sin^2\thetanu_{23}$
than the \textbf{LO} solution, consistent with approximate
Eqs.~\eqref{eq:th13nu_LO_13b}--\eqref{eq:th23nu_LO_13} which, for this rotation, hold for any $\thetal_{12}$.
By contrast, $\sin^2\thetanu_{12}$ exhibits a broad marginal distribution and the
contours are clearly elongated along this direction, reflecting the large spread
introduced by scanning $\thetal_{12}$. For this angle, the peak of the marginal distribution is located towards $\thetanu_{12}\to0$; this result is consistent with that shown in Fig. \ref{fig:angles_vs_theta_12} where fixing all the input parameter to their best fits this angle is always the smallest one. However, the tail of the distribution can reach $\sin^2\thetanu_{12}\sim0.2$ for both octant cases.

\subsection{Case II: (1,3) charged-lepton rotation}
\label{sec:results_13}

\begin{figure}[t]
    \centering
    \includegraphics[width=0.7\linewidth]{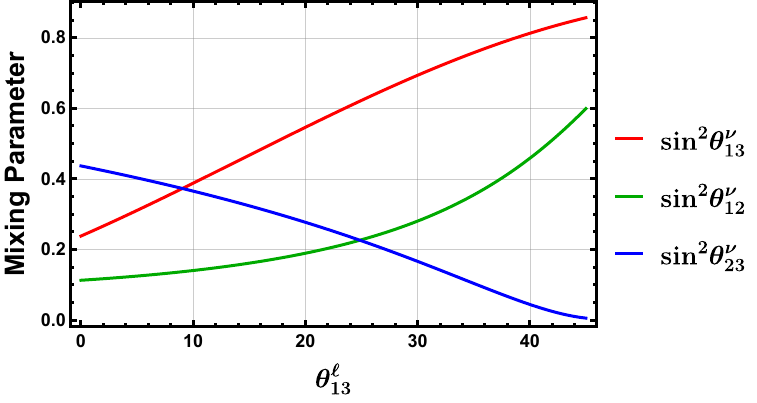}
    \caption{Same as Fig.~\ref{fig:angles_vs_theta_12} but for the (1,3) charged-lepton
    rotation, as a function of $\thetal_{13}$.}
    \label{fig:angles_vs_theta_13}
\end{figure}

Figure~\ref{fig:angles_vs_theta_13} shows the the values of neutrino angles obtained for various
$\thetal_{13}$. In striking contrast to Case~I, all the angles drastically change as $\thetal_{13}$ varies, in agreement
with the analytic predictions. While ${\sin^2\thetanu_{12}}$ and ${\sin^2\thetanu_{13}}$ increase to very large value as $\thetal_{13}$ increases, $\sin^2\thetanu_{23}$ sharply decreases toward 0 for $\thetal_{13}\to45^\circ$. The hierarchy $\thetanu_{23}>\thetanu_{13},\thetanu_{12}$ between the neutrino mixing angle for $\thetal_{13}\to0$ shown in the analytic section (see Eqs.~\eqref{eq:th13nu_LO_13b}-\eqref{eq:th12nu_LO_13}) is well observed but only holds for $\thetal_{13}<10^\circ$.

\begin{figure}[tbp]
    \centering
    \includegraphics[width=0.95\linewidth]{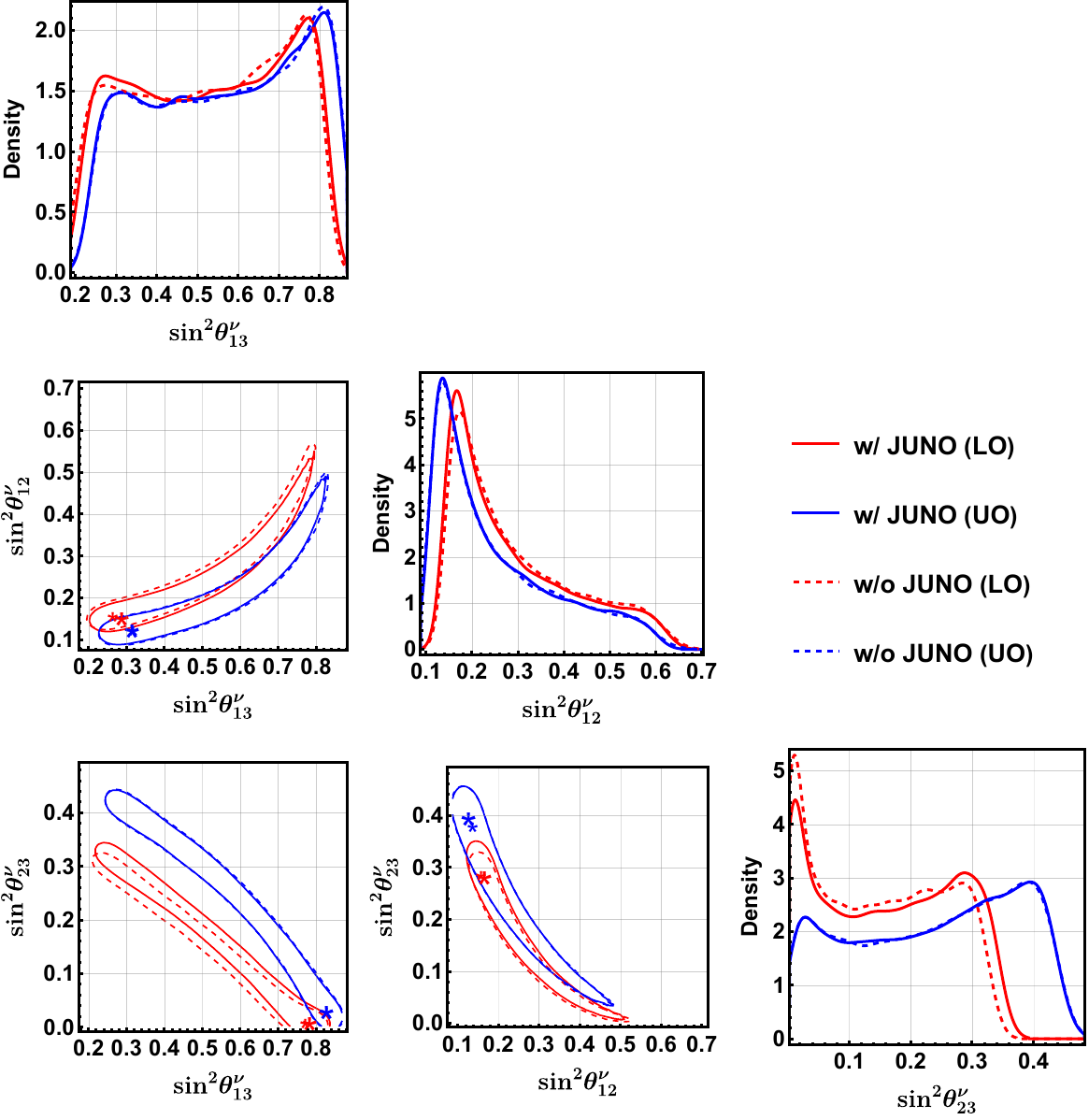}
    \caption{Same as Fig.~\ref{fig:triangle12} but for the (1,3) single charged-lepton
    rotation case.}
    \label{fig:triangle13}
\end{figure}

Figure~\ref{fig:triangle13} shows the marginal distributions of the three
neutrino mixing angles together with the 90\% contours in the three pairs of neutrino
angle planes, obtained with the same procedure described for Case~I.
The shapes are noticeably different from Case~I: in this case all the three neutrino mixing angle cannot be determined independently from $\thetal_{13}$. Therefore, despite being peaked at some specific values, all the three marginal distributions are considerably broad. It is worth mentioning that the $\theta_{23}$ octant is extremely important for the $\thetanu_{23}$ distribution, for which the preferred value is much lower in the \textbf{LO} case than in the \textbf{UO} case. However, for all the angles, the \textbf{LO} and \textbf{UO} regions (red and blue) are not clearly separated.
Another distinctive feature of this case is the strong correlations among the three angles, in particular positive between $\sin^2\thetanu_{12}$ and $\sin^2\thetanu_{13}$ and negative for the other two angle combinations.
We have also checked that fo small $\thetal_{13}$, the correlation between $\sin^2\thetanu_{12}$ and $\sin^2\thetanu_{23}$ becomes strongly positive, reflecting the analytical relation valid for $\thetal_{13}\to0$, 
$\sin^2\thetanu_{12}/\sin^2\thetanu_{23}\sim\sin^2\theta_{12}\cot^2\theta_{23}$.

The impact of JUNO is visible, but not drastic. Indeed, since the fixed second
column depends on $\theta_{12}$ and $\theta_{23}$ through the combination
$\sin\theta_{12}\cos\theta_{23}$, the improvement in $\theta_{12}$ precision leads
only to a slight narrowing of the marginal distributions, being the atmospheric mixing angle uncertainty still dominating. Future precision data from DUNE and T2HK narrowing the uncertainty on $\theta_{23}$ and determining its octant
will be therefore the decisive probe of Case~II.

\subsection{Case III: (2,3) charged-lepton rotation}
\label{sec:results_23}

\begin{figure}[t]
    \centering
    \includegraphics[width=0.7\linewidth]{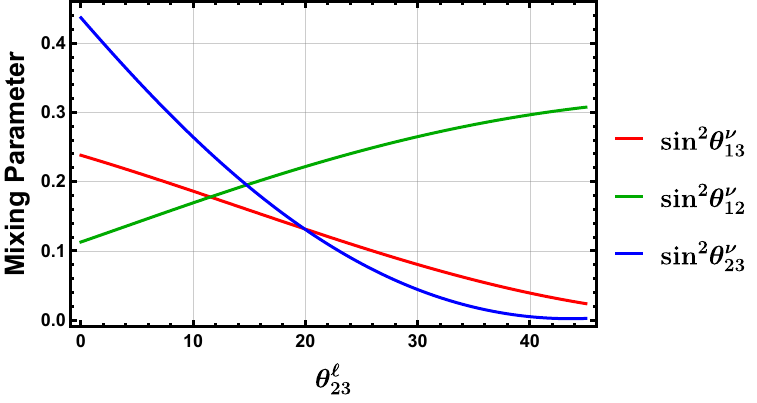}
    \caption{Same as Fig.~\ref{fig:angles_vs_theta_12} but for the (2,3) charged-lepton
    rotation, as a function of $\thetal_{23}$.}
    \label{fig:angles_vs_theta_23}
\end{figure}

Figure~\ref{fig:angles_vs_theta_23} illustrates the dependence of the neutrino angles
on $\thetal_{23}$. The solar angle $\sin^2\thetanu_{12}$ (green) increases across the full range of $\thetal_{23}$.  The reactor angle $\sin^2\thetanu_{13}$ (red)
decreases from its initial value $\sin^2\theta_{12}\sin^2\theta_{23}\approx 0.17$
at $\thetal_{23}=0$, as predicted by Eq.~\eqref{eq:th13nu_case3_LO} with the effective
angle $\Delta_{23}=\theta_{23}-\thetal_{23}$ decreasing with $\thetal_{23}$. The
atmospheric angle $\sin^2\thetanu_{23}$ (blue) decreases steeply as $\thetal_{23}$
increases, following the sum rule~\eqref{eq:sumrule_23}: at $\thetal_{23}=\theta_{23}$
(where $\Delta_{23}=0$) both $\sin^2\thetanu_{13}$ and $\sin^2\thetanu_{23}$ would
vanish, leaving only $\thetanu_{12}$.

\begin{figure}[htbp]
    \centering
    \includegraphics[width=0.95\linewidth]{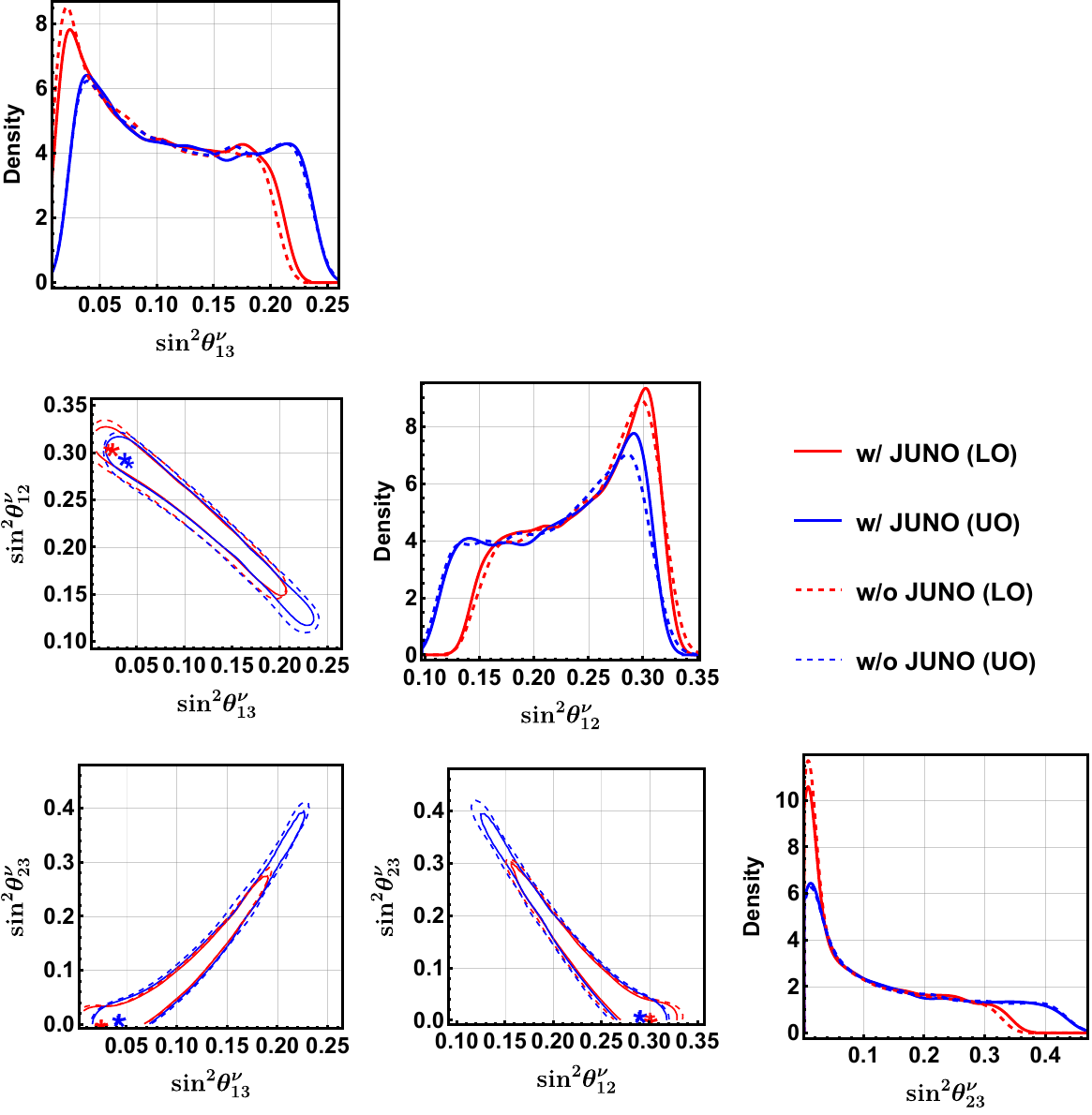}
    \caption{Same as Fig.~\ref{fig:triangle12} but for the (2,3) charged-lepton
    rotation case.}
    \label{fig:triangle23}
\end{figure}

Figure~\ref{fig:triangle23} shows the marginal distributions of the three
neutrino mixing angles together with the 90\% contours in the three pairs of neutrino
angle planes, obtained with the same procedure described for Cases~I and~II.
The most distinctive feature is the negative correlation between $\sin^2\thetanu_{13}$
and $\sin^2\thetanu_{12}$, confirmed by the exact
prediction in~\eqref{eq:sumrule_case3_exact}. The marginal distributions of all the three angles are broad, reflecting their strong dependence by the
effective combination $\Delta_{23} = \theta_{23} - \thetal_{23}$.
The \textbf{LO} and \textbf{UO} solutions (red and blue) are not well separated, with the distribution being broader in the \textbf{UO} case.
The impact of JUNO is mild for $\thetanu_{13}$ and $\thetanu_{23}$, whose spread
is dominated by the scan over $\thetal_{23}$, while a slight improvement is visible
in the $\thetanu_{12}$ distribution, with the NuFit~6.1 contours (solid) being
somewhat smaller than the NuFit~6.0 ones (dashed) in all planes.

\section{Effect of the CP-violating phase \boldmath{$\delta_{\rm CP}$}}
\label{sec:deltaCP}

The analytical results derived in Section~\ref{ssec:analytical} were mostly obtained 
in the limit $\theta_{13}\to 0$, which---while transparent and compact---entirely
suppresses the dependence on the Dirac CP phase $\delta_{\rm CP}$.
Since $s_{13} \equiv \sin\theta_{13} \approx 0.149$ is not negligible at the
current level of precision, it is important to assess how $\delta_{\rm CP}$ modifies
the predictions for the neutrino mixing angles $\thetanu_{ij}$ once the reactor angle is
restored to its measured value (see \cite{Girardi:2015vha} for a systematic review on this topic).

\begin{figure}[tbp]
  \centering
  \includegraphics[width=\linewidth]{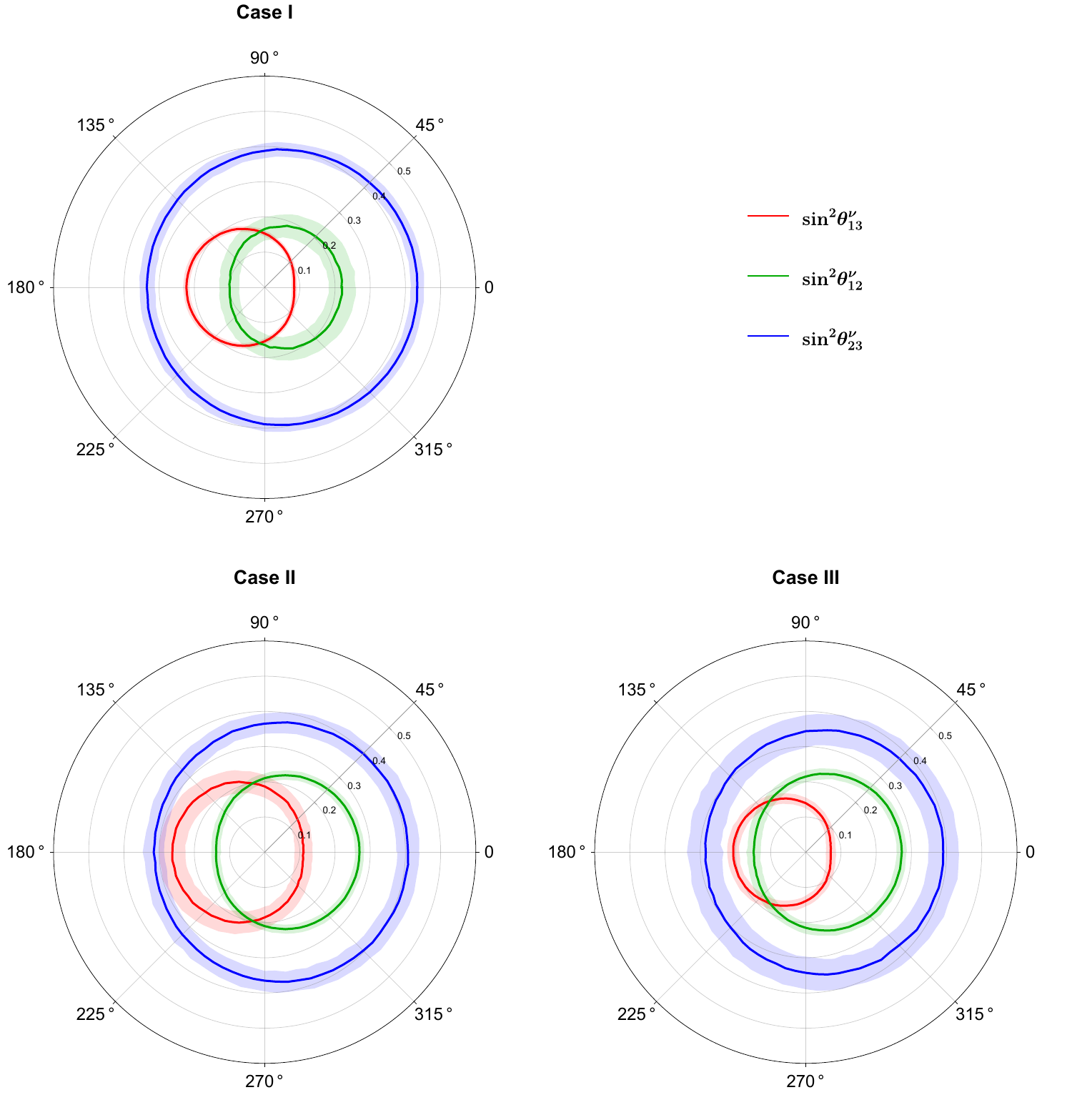}
  \caption{Polar plots of the neutrino mixing angles $\sin^2\thetanu_{13}$ (red),
  $\sin^2\thetanu_{12}$ (green), and $\sin^2\thetanu_{23}$ (blue) as a function
  of $\delta_{\rm CP}\in[0^\circ,360^\circ)$, shown for the three charged-lepton
  rotation scenarios (Cases~I, II, III). The radial coordinate
  gives the value of each $\sin^2\thetanu_{ij}$, and the shaded bands indicate the
  $1\sigma$ uncertainty from the PMNS input parameters (NuFit~6.1, \textbf{NO}, \textbf{UO}).
  The leptonic rotation angle is constrained to $\thetal_{ij}<6^\circ$.
  The eccentricity of each orbit reflects the magnitude of the $\cos\delta_{\rm CP}$
  modulation at order $\sin\theta_{13}$, while the mean radius mainly encodes the leading-order
  (i.e.\ $\theta_{13}=0$) prediction.}
  \label{fig:radar_deltaCP}
\end{figure}

Figure~\ref{fig:radar_deltaCP} displays the values of $\sin^2\thetanu_{13}$ (red),
$\sin^2\thetanu_{12}$ (green), and $\sin^2\thetanu_{23}$ (blue) as polar curves in the
$\delta_{\rm CP}$ plane, with the angular coordinate running over $\delta_{\rm CP}\in[0^\circ,360^\circ)$.
The radial distance from the origin encodes the value of each mixing angle squared.
The shaded bands indicate the $1\sigma$ uncertainty propagated from the PMNS input
parameters, with the leptonic rotation angle constrained to $\sin\thetal_{ij}<0.1$ to suppress the mixing angle spreads caused by its variance. The three panels correspond to
Cases I, II, and III, respectively. 
Fig.~\ref{fig:radar_deltaCP} provide a compact visual summary of the $\delta_{\rm{CP}}$ dependence of the neutrino mixing angles: the
eccentricity of each orbit measures the $\delta_{\rm CP}$ sensitivity of the
corresponding angle, while its mean radial distance encodes the leading-order
prediction (i.e. $\theta_{13}\to0$). As the measurement of $\delta_{\rm CP}$ improves with DUNE and T2HK, the
CP-phase modulation captured in Fig.~\ref{fig:radar_deltaCP} will become a direct
quantitative test of the rotation hypothesis assumed for $\Ul$, at least in the case of small leptonic mixing angle.


In the \textbf{Case~I}, the key analytical result is Eq.~\eqref{eq:th13nu_LO_12}, which
already captures the full $\delta_{\rm CP}$ dependence of $\sin^2\thetanu_{13}$ at
order $\sin\theta_{13}$:
\begin{equation}
  \sin^2\thetanu_{13}
  \approx \sin^2\theta_{12}\sin^2\theta_{23}
    - \tfrac{1}{2}\sin 2\theta_{12}\,\sin 2\theta_{23}\sin\theta_{13}\cos\delta\,,
\end{equation}
where the last term is the only source of $\delta_{\rm CP}$ dependence. It modulates
$\sin^2\thetanu_{13}$ sinusoidally, with amplitude
\begin{equation}
  \mathcal{A}_{13}^{(12)}
  \equiv \tfrac{1}{2}\sin 2\theta_{12}\,\sin 2\theta_{23}\sin\theta_{13}
  \approx 0.07\,,
  \label{eq:amplitude_case12}
\end{equation}
using the NuFit~6.1 best-fit values. This amplitude represents
roughly $40\%$ of the leading-order term $\sin^2\theta_{12}\sin^2\theta_{23}\approx 0.17$
(\textbf{UO}), so that $\sin^2\thetanu_{13}$ ranges from approximately $0.10$ at
$\delta_{\rm CP}=0^\circ$ to $0.24$ at $\delta_{\rm CP}=180^\circ$.
This small $\cos\delta_{\rm{CP}}$ modulation is visible in the first panel of
Fig.~\ref{fig:radar_deltaCP} as the distinctly slightly non-circular red orbit.

The atmospheric angle $\sin^2\thetanu_{23}$ is, at leading order in $\sin\theta_{13}$,
predicted by Eq.~\eqref{eq:th23nu_LO_13} and it is independent of both $\thetal_{12}$
and $\delta_{\rm CP}$. The correction at order $\sin\theta_{13}$ arises from the $(2,3)$ entry
of the third column of $\Unu^{(12)}$:
\begin{align}
  |(\Unu^{(12)})_{23}|^2
  &= \cos\theta_{12}^2 \sin\theta_{23}^2 + \sin\theta_{12}^2 \cos\theta_{23}^2 s_{13}^2
     + \tfrac{1}{2}\sin 2\theta_{12}\,\sin 2\theta_{23}\sin\theta_{13}\cos\delta\,,
\end{align}
which, combined with the denominator $1 - \sin^2\thetanu_{13}$, gives a
$\delta_{\rm CP}$ correction to $\sin^2\theta_{23}^\nu$ of order
\begin{equation}
\dfrac{\sin 2\theta_{12} \sin2\theta_{23}\cos^2\theta_{23}}{2(1-\sin^2\theta_{12}\sin^2\theta_{23})^2}\approx0.04\,,
\end{equation}
smaller than the correction for the $\sin^2\theta_{13}^\nu$.
 The blue orbit in
Fig.~\ref{fig:radar_deltaCP} (Case~I panel) is therefore almost circular. The solar angle $\sin^2\thetanu_{12}$ (green), which depends on
$\thetal_{12}$ through the first two columns of $\Unu^{(12)}$, also acquires
an $\mathcal{O}(\sin\theta_{13})$ correction proportional to $\cos\delta_{\rm{CP}}$; its orbit
is intermediate in eccentricity between those of the other two angles.

The second panel of
Fig.~\ref{fig:radar_deltaCP} shows the same pattern for the \textbf{Case~II}. This can be understood from the analytical expression of the second column of $U_\nu^{(13)}$.
The $\delta_{\rm CP}$ dependence enters
through the $\mathcal{O}(\sin\theta_{13})$ corrections
as
\begin{equation}
    \begin{aligned}
  (\Unu^{(13)})_{12} &= -\sin\theta_{12}\cos\theta_{23} - \cos\theta_{12}\sin\theta_{23}\,\sin\theta_{13}\,e^{-i\delta}\,,\\
  (\Unu^{(13)})_{22} &= \cos\theta_{12}\cos\theta_{23} - \sin\theta_{12}\sin\theta_{23}\,\sin\theta_{13}\,e^{-i\delta}\,,
\end{aligned}
\end{equation}
Taking moduli squared and keeping terms linear in $\sin\theta_{13}$,
both of these entries acquire a $\cos\delta_{\rm{CP}}$ correction with the same overall
amplitude $\mathcal{A}^{(13)} = \frac{1}{2}\sin 2\theta_{12}\,\sin 2\theta_{23}\sin\theta_{13}$, just
as in Case~I. 


\textbf{Case~III} displays a qualitatively different pattern. The exact sum-rule
Eq.~\eqref{eq:sumrule_case3_exact} reads
$\cos^2\thetanu_{12}\cos^2\thetanu_{13} = \cos^2\theta_{12}\cos^2\theta_{13}$
and is valid for \textit{any} value of $\thetal_{23}$ and $\delta_{\rm CP}$: it is genuinely
phase-independent. The reason is structural: the first column of $\Unu^{(23)}$ is
identical to the first column of $\UPMNS^\dagger$, namely
$(\cos\theta_{12}\cos\theta_{13},\, \sin\theta_{12}\cos\theta_{13},\, \sin\theta_{13}e^{i\delta})^T$. Since mixing angles
are extracted from the moduli of matrix elements, the entire first column---and hence the exact constraint
Eq.~\eqref{eq:sumrule_case3_exact}---is independent of $\delta_{\rm CP}$ to all orders in
$\theta_{13}$. This is reflected in the right panel in the second row of Fig.~\ref{fig:radar_deltaCP}:
the green orbit ($\sin^2\thetanu_{12}$) is the most nearly circular, consistent with its $\delta_{\rm{CP}}$-independent determination
at leading order. This panel also shows that $\sin^2\thetanu_{13}$ (red) varies the most with $\delta_{\rm CP}$ carrying $\mathcal{O}(\sin\theta_{13})$
phase-sensitive entries analogous to those discussed in Cases~I and~II.



\section{The Jarlskog invariant of \boldmath{$\Unu$}}
\label{sec:jarlskog}

\begin{figure}[tbp]
  \centering
\includegraphics[width=\linewidth]{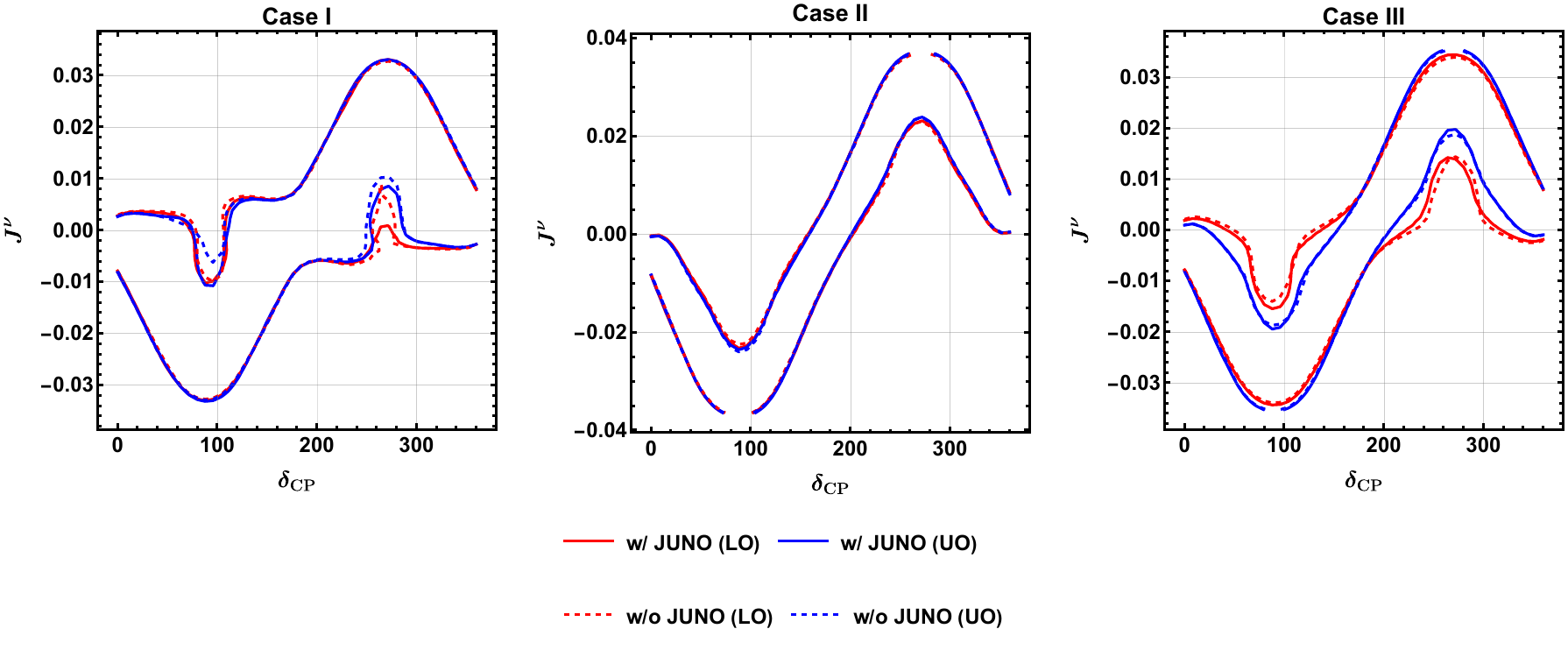}
  \caption{90\% contour regions of the Jarlskog invariant $J^\nu$ of the neutrino
  mixing matrix as a function of $\delta_{\rm CP}$, for Cases~I, II, and~III
  (left, center, right). Solid (dashed) curves correspond to NuFit~6.1 (NuFit~6.0)
  inputs; red (blue) curves refer to the \textbf{LO} (\textbf{UO}) solution of $\theta_{23}$.
  The CP phase $\delta_{\rm CP}$ is scanned uniformly over $[0^\circ,360^\circ)$.}
  \label{fig:jarlskog}
\end{figure}

A probe of the CP structure of $\Unu$ is provided by its Jarlskog
invariant~\cite{Jarlskog:1985ht}, defined as
\begin{equation}
  J^\nu \equiv \mathrm{Im}\!\left[
    (U_\nu)_{e1}\,(U_\nu)_{\mu 2}\,
    (U_\nu)_{e2}^*\,(U_\nu)_{\mu 1}^*
  \right]
  = \frac{1}{8}
    \sin 2\thetanu_{12}\,\sin 2\thetanu_{13}\,\sin 2\thetanu_{23}\,
    \cos\thetanu_{13}\,\sin\delta^\nu\,,
  \label{eq:jarlskog_def}
\end{equation}
where the second equality holds once $\Unu$ is brought to the PDG
parametrization.  Compared to the neutrino mixing angles studied in
Section~\ref{sec:results}, $J^\nu$ encodes the full interplay of all three
mixing angles \emph{and} the CP phase of the neutrino sector, making it a
particularly sensitive probe of the rotation hypothesis assumed for $\Ul$.
We extract $J^\nu$ directly from the matrix $\Unu = \UPMNS^\dagger\,\Ul$ using
the first equality in Eq.~\eqref{eq:jarlskog_def}, without rephasing, following
exactly the same procedure described in Section~\ref{sec:results}.
Since $\delta_{\rm CP}$ is now a dynamical variable of the scan rather than a
fixed input, each point corresponds to a pair $(\delta_{\rm CP},\,J^\nu)$; we display the resulting 90\% density contours in the
$(\delta_{\rm CP},\,J^\nu)$ plane. The results for the three rotation cases are
shown in Fig.~\ref{fig:jarlskog}.
In the limit of small charged-lepton mixing angles $\theta^l_{ij}$, the Jarlskog invariant scales as $J\propto \sin\delta_{CP}$. This sinusoidal modulation represents the dominant feature in all the three panels in Fig.~\ref{fig:jarlskog}, and it is a model-independent consequence of the PDG parametrization applied to
$\Unu$.
The dominant feature in all three panels is the characteristic sinusoidal modulation
$J^\nu \propto \sin\delta_{\rm CP}$, which follows directly from Eq.~\eqref{eq:jarlskog_def}
and is a model-independent consequence of the PDG parametrization applied to
$\Unu$. The amplitude of this modulation is set by the product of the three neutrino
mixing angles, which in turn are functions of the PMNS angles evaluated at
$\thetal_{ij} = 0$. At leading order in $\theta_{13}$ and for small
$\thetal_{ij}$, combining Eqs.~\eqref{eq:th13nu_LO_13b}--\eqref{eq:th12nu_LO_13}
with Eq.~\eqref{eq:jarlskog_def} 
we obtain a maximum amplitude $|J^\nu|_{\rm max} \approx 0.03$ in good agreement with the figure.

The three cases share the same qualitative sinusoidal shape, but differ in the
width of the 90\% band and in the octant separation. In particular, the bands result to be wider in Case I, due to the lack of correlation between neutrino mixing angles. The effect of octant is more pronounced in Case III, where the distribution of the possible neutrino mixing angles are always larger in the \textbf{UO} case. 
Another distinctive feature visible in Fig.~\ref{fig:jarlskog} is the narrowing of the
90\% bands near maximal CP violation, $\delta_{\rm CP} \simeq \pm 90^\circ$, most
prominently in Cases~I and~III (\textbf{LO}). This can be understood analytically:
since $J^\nu \propto A(\thetal)\sin(\delta_{\rm CP}+\varphi(\thetal))$, the spread
induced by scanning $\thetal$ receives two contributions, proportional to $\sin\delta_{\rm CP}$
(amplitude variation $\Delta A$) and $\cos\delta_{\rm CP}$ (phase shift $\Delta\varphi$),
respectively. At $\delta_{\rm CP}=\pm90^\circ$ the
$\cos\delta_{\rm CP}$ term vanishes, so only $\Delta A$ contributes. In Cases~I
and~III the neutrino mixing amplitudes are less sensitive to $\thetal$ (see Figs \ref{fig:angles_vs_theta_12} and \ref{fig:angles_vs_theta_23}); thus,
$\Delta A$ is small and the band narrows. In Case~II, by contrast, all three neutrino
angles vary strongly with $\thetal_{13}$ (see Fig. \ref{fig:angles_vs_theta_13}), making $\Delta A$ large enough to keep the
band wide throughout the full $\delta_{\rm CP}$ range.
Finally, we observe how the JUNO measurement only have a marginal impact in this analysis; indeed,
only future measurements by
reducing the uncertainty on the atmospheric angle will
provide significant improvement in constraining $J^\nu$ in all three
scenarios.

\section{Conclusions}
\label{sec:conclusions}

We have analyzed the constraints on the neutrino mixing matrix $\Unu = \UPMNS^\dagger\,\Ul$
arising from the hypothesis that the charged-lepton mixing matrix $\Ul$ is a single
rotation in the (1,2), (1,3), or (2,3) sector. Our main findings can be summarized
as follows.

In each case, one column of $\Unu$ is entirely fixed by the PMNS parameters and
independent of the charged-lepton rotation angle $\thetal$. This structural observation
leads to sharp, parameter-free predictions for certain combinations of neutrino mixing
angles: in Case~I the neutrino reactor and atmospheric angles are predicted from the
PMNS solar and atmospheric angles alone (Eqs.~\eqref{eq:th13nu_LO_13b}--\eqref{eq:th23nu_LO_13}),
with the solar angle obeying the generalized sum rule~\eqref{eq:th12nu_sumrule_12};
in Case~II, the ratio in Eq.~\eqref{eq:caseIIhierarchy} fixes the hierarchy among the neutrino mixing angles;
in Case~III, the exact relation $\cos^2\thetanu_{12}\cos^2\thetanu_{13}=\cos^2\theta_{12}\cos^2\theta_{13}$
holds for any value of $\thetal_{23}$ (Eq.~\eqref{eq:sumrule_case3_exact}), while
the atmospheric angle follows the compact sum rule $\thetanu_{23}\approx\theta_{23}-\thetal_{23}$.

The numerical results confirm and sharpen these analytical findings. In Case~I,
the JUNO measurement of $\theta_{12}$ directly narrows the predicted ranges of
$\sin^2\thetanu_{13}$ and $\sin^2\thetanu_{23}$, as both depends by $\theta_{12}$ at leading order. The improvement is clearly
visible in the tightening of the post-JUNO contours (Fig.~\ref{fig:triangle12}). In Case~II, the dominant uncertainties come from
$\theta_{23}$ and the scan over $\thetal_{13}$, and the impact of JUNO is more
modest; future precision measurements of $\theta_{23}$ by DUNE and T2HK will be
the decisive probe of this scenario. In Case~III, the exact analytical constraint
manifests as a negative correlation in the $(\sin^2\thetanu_{12},\sin^2\thetanu_{13})$ plane,
a distinctive and testable signature of the (2,3) rotation hypothesis.
We also checked the impact of the $\delta_{\rm{CP}}$ uncertainty on the determination of the neutrino mixing angle,  as well as the possible values of the Jarlskog invariant of the neutrino matrix. 

Taken together, our results demonstrate that even under the simple hypothesis
of a single-sector charged-lepton rotation, the current precision on the PMNS
mixing angles already allows one to set meaningful constraints on the structure
of the neutrino mixing matrix $\Unu$, providing a concrete and testable framework
in the context of flavor physics model building. We have also shown that the first
JUNO measurement of $\theta_{12}$, despite being based on a limited dataset, leads
to a noticeable (even though not drastic) improvement in the predicted ranges of the neutrino mixing angles, underscoring the importance of achieving sub-percent
precision on all oscillation parameters. As the experimental program advances
with DUNE, T2HK, and the full JUNO dataset, the tests discussed here---and their
natural extensions to other rotation hypotheses---will become increasingly
definitive probes of the underlying symmetry structure of the lepton sector.

\acknowledgments
SM acknowledges financial support from the project \emph{CPI–25–643}, part of the \emph{CNS2022-136005}, founded by \emph{MICIU/AEI/10.13039/501100011033} for the \emph{Uni\'on Europea NextGenerati\'onEU/PRTR}.
\flushbottom
\newpage
\bibliographystyle{JHEP}
\bibliography{bibliography}
\end{document}